\documentclass{article}
  \usepackage{graphicx}
  \usepackage{authblk}
\usepackage{ragged2e}
\usepackage{amsmath}
\usepackage{abstract}

\begin{document}
\title{\bf The Smoke of Zaanen}
\author{Philip W. Phillips}
\date{}
\affil{Department of Physics and Anthony J. Leggett Institute of Condensed Matter Theory, University of Illinois at Urbana-Champaign, Urbana, IL 61801, USA}

\maketitle
  
\begin{abstract}
\begin{center}
    Theoretical physics suffered a major loss with the death of my dear friend Jan Zaanen on January 18.  This note is my remembrance of him.  
\end{center}
\end{abstract}

\begin{figure}[h!]
    \centering\includegraphics[width=0.45\textwidth]{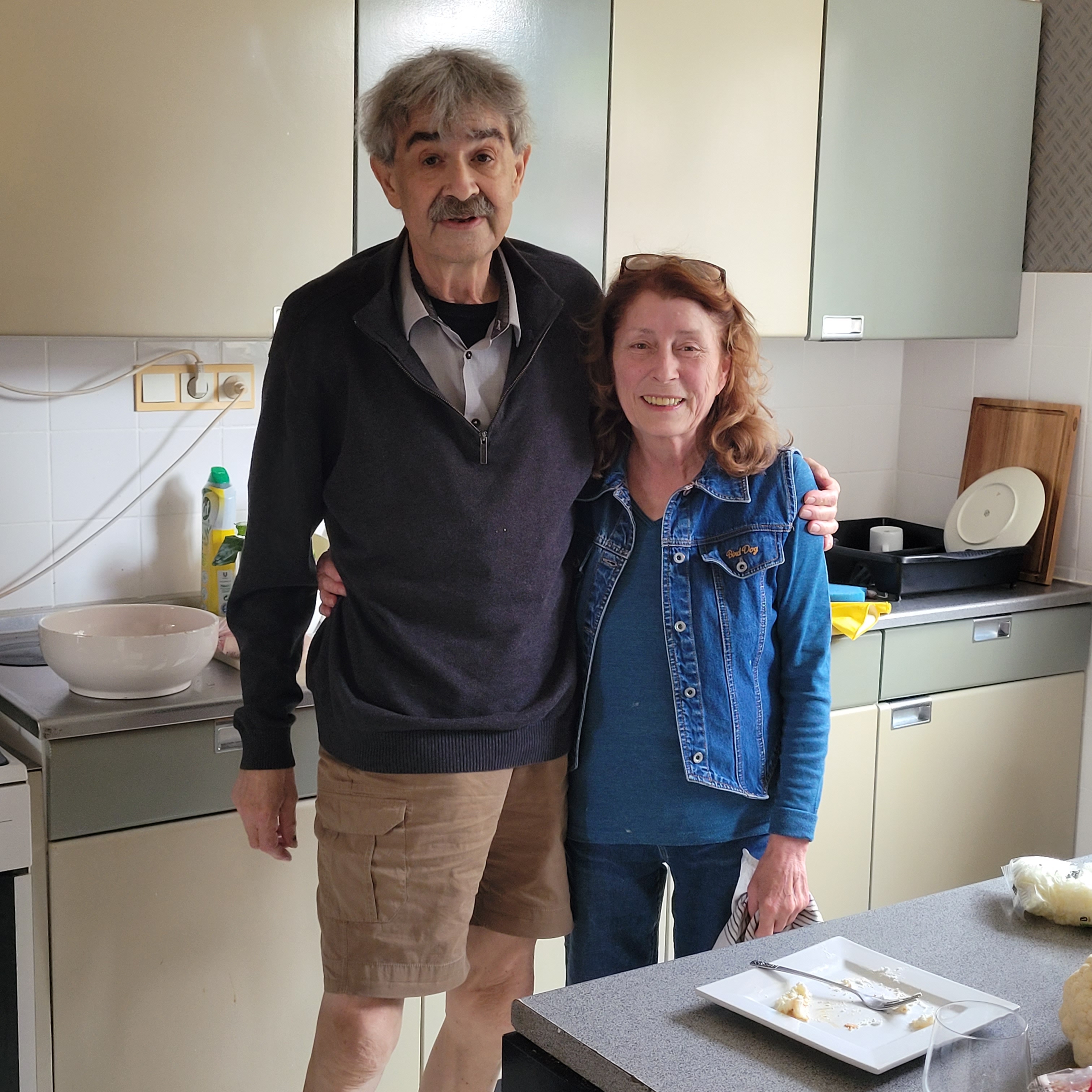}
    \caption{Jan and Christa: July 18, 2023, Leiden}
\end{figure}
  
Johannes (Jan) Zaanen, a resonant, much-needed, deftly insightful voice in  our field of strongly correlated quantum matter, died Jan. 18 2024 in Leiden, the place of his birth.   He was 66.  I received the news from Erik van Heumen, a colleague of his at U. Amsterdam who was handling most of Jan's scientific and public matters once he took ill.  In fact,  they had just co-written a paper, which they submitted to Science, on the breakdown of spectral sum rules in the underdoped regime of  the copper-oxide high-temperature superconductors.  This was Jan's favorite problem.   He threw the kitchen sink at it always looking for something new.  His methodology ranged from more traditional, early Hartree-Fock\cite{hf} studies on stripes in the cuprates ( though it was highly influential, he never recounted it fondly), the local density approximation augmented with the Hubbard $U$\cite{ldau} (LDA+U) to his more exotic mining\cite{ads} of the gauge gravity duality to gain traction on the strange metal T-linear resistivity problem in the cuprates.  As we had both become gauge-gravity duality converts, we would attend joint meetings with string theorists and discuss how to really push this technology to realistic finite-density matter.    In fact, in our last Zoom call on Dec. 30, 2023,  (a meeting for which I was an hour late because of a time-zone mixup, Jan entreating me ``We had a date, i kept hands free and now i am waiting in vain for you, not great'') ,  the old problem of Sawatzky's\cite{sawa}  dynamical spectral weight transfer came up as a possible route to strange metallicity.   To the end, the strange metal remained his burning light.  We had agreed to write a joint paper on it putting all the details of Mottness into the mix.  Along with physics, Jan laid plain his frailty.   Talking to me from his hospice bed, he stated what I could see all too clearly:  ``I am basically paralyzed you know and the mind is the last to go.''     He talked about empathy and ``humankind being pretty nice.''  He continued, ``You know Philip, I am back to being human.  Physics keeps me going but now I'm human.''

Obituaries are unwieldy because of the obvious weight that goes with writing them.  What is difficult to communicate is a snapshot into the person's actual life, how it was like for them, not just a global summary.  It is for this reason that I have decided to bring Jan to the fore with a story.  This story, `The smoke of Zaanen,' is a love story.  Not a typical one though. Most love stories are private, lots of intimate exchanges, long looks and commitments that last.  But this one took place out in the open, typically in the front of buildings.  We all saw it happen, except the beginning.  I had heard the story before but never thought much of it.  Jan always tells stories  but listening to him on March 21, 2023 in the hospital  as fluid was being drained from his lungs made his words more paramount. Human frailty focuses the empathic mind.  Jan was suffering from complications from two surgeries to relieve esophageal cancer.
 
When I saw Jan in July (2022) in Stockholm, there was no hint of hospitalization in his future.  At that meeting, he chose the Swedish means of transportation, the stand-up two-wheel slightly motorized scooter, to get around Stockholm.  He seemed like a child again, each morning regaling me and others within earshot of the joy of 2-wheel standing mobility around Stockholm.  I settled for the bus, the lack of European cell service being the bridge too far, or more aptly the expense I was not willing to incur.  I saw him take off one morning with others from the workshop at Nordita and indeed he seemed a natural at it, no reticence there.  I suspected he must have ridden a rudimentary version of the modernized Segway transport when he was young.   It turns out Jan and I spent part of our childhoods on the most southern  islands  in the Caribbean, Jan in Curucao and Tobago for me.  Jan was conceived on a family vacation to Costa Rica.  After a brief stint back in Curucao, he was shuttled to Leiden where he entered the world in the very hospital in which I was then visiting him, Leiden University Medical Center.  Immediately after, they returned to Curucao for the next three and a half years giving  shape to his  Caribbean roots.  Perhaps this is part of our mutual bond, he intonated.

Back to March 2023.  In between nurses' visits , we continued our conversation. I noticed the level of the fluid  in the drain container was rising.    Jan insisted it wasn't and continued, doing things his way, another of Jan's traits.   As a goodbye  to the Caribbean,  Jan's parents took a trip to Mexico.  Jan's father, a mechanical engineer for Shell, was on kid duty that day in the open-air market in Mexico City as his mother was taking a break.  Wandering through the market with his sister, Mieke, born in Curucao, he noticed a quite irresistible lump on the ground which he proceeded to eat.  Much to his father's surprise, it was horse shit.  Yes, Jan ate horse shit.  The cure for the dysentery that ensued was a strict diet of ripe bananas.  How they managed to smuggle enough bananas on their  plane carrying them back to the Netherlands remains a mystery, a story shrouded in Dutch customs lore:  the Dutch family with a year's supply of bananas and an extremely hungry kid. 

Like all of us who grew up in the Caribbean, winter is harsh.   Hearing a knock on the door one day, he opened it and was greeted with what seemed like a fierce animal.  It was the chill of winter, his first.  Nothing in Curucao readied him to be pierced  so unceremoniously with just the opening of a door.   There was no warning,  just an enduring cold.  Throughout his life, Jan hated winter,  thoughts back to Curucao his only escape.

Around age 13,  Jan discovered another escape, and here lies the love story.  One day his school had a class outing to the countryside in Drenthe just outside Groningen.    Two of his class mates, Marianna and Maartje took him on a hike away from the crowd.  Jan was a good-looking boy but shy and nerdy so there was no resistance to being led by two precocious girls, also 13.  They tired and soon spotted a park bench.  I know what you're thinking but that's not it.  This is a PG story about a nerdy school boy who would become a world-renowned physicist,  coiner of Planckian dissipation\cite{planckian} and the classifier\cite{zsa} of band gaps in transition metal compounds which led to his 1985 thesis with the man who ``showed him physics,''  George Sawatzky.     

 There is of course continuity in how we become who we are.      Or perhaps not?   But we'll see. The girls, placing Jan firmly on the bench, said,  ``we have something for you.''    Jan liked Maartje but was too shy to say so.     Marianna sat on his right knee and Maartje on his left. Jan heard some undecipherable rustling but he kept his  eyes closed as instructed.    Was this the moment in this nerd's life?   In a flash Jan envisioned his future as a rock and roller guitarist (something he did try before grad school)--perfect training at hand.   Ever so carefully, Marianna took out wrapping paper and tobacco and rolled a cigarette and placed it in between  Jan's lips.  Maartje took out a lighter and lit the cigarette.   He then opened his eyes as Maartje clasped his hand and led it to meet the cigarette.   What an introduction to smoking.  ``You know Philip, the first 4 cigarettes are pretty bad,  but then the fifth one, that's an experience,'' Jan said.  They returned to the group, Jan now in love.  He discovered smoking and it would forever be linked to Marianna and Maartje, each puff taking him back to that day in Drenthe  where his world changed to the Jan Zaanen we knew.  
     
 I first met Jan in 1997/98 in Trieste and of course it was on one of his notorious smoking breaks, complete with the hypnotic rolling of tobacco.  He stopped me and said he liked my talk (on the Kravchenko problem as I recall) and proceeded to say some things I barely followed.  Jan's accent took a while to decipher but his love for smoking was as clear as a crystal day.  When I saw him smoking in Trieste, I remember thinking why doesn't this man just buy the damn cigarettes in a store.   With the back story revealed, I now see the continuity.  I now look back on that first meeting in Trieste differently.  There was a certain intensity and earnestness in Jan's eyes when he smoked.  It was as if something were being revealed to him.  I now have an idea what.  From the first cigarette Marianna rolled for him ever so deliberately,  smoking  would always take him back to that park bench in Drenthe when he was 13.  He remained purely platonic with Maartje, and of course Marianna, but their mark on him was indelible.  
  
  Jan was not a man of few words.  He was voluble and wide ranging in his views on physics.  I hung out with him because we had similar likes and dislikes.  He liked children, my son no exception as we spent much time together in Kyoto.  
  We both shared a dislike for conventional approaches to correlated electron problems.  So strong was Jan's drive to do something new that he turned down a tenured job at Stuttgart for a post-doc at Bell Labs in 1987.  For him, Bell Labs ``was playing on Mt Olympus.''  So why not.  He scaled those heights and was rewarded with the position he filled throughout the remainder of his life, Professor of physics at Lorentz' institution, University of Leiden.  He was a Spinoza Award recipient, the highest honor in the Netherlands in science as well as a member of the Dutch Royal Academy.   In July 2023, many of us convened in Leiden for a celebration of Jan's remarkable life in physics.  Of course the meeting was organized by Erik van Heumen.  We all paid our debt to Jan for inspiring us to think differently.   Undoubtedly, his non-traditional background as a chemist as an undergraduate prepared him to think outside the box in theoretical physics.  This commonality in our backgrounds cemented the bond between us.
    
  The day after the meeting, I rode a bike out to see Jan and Christa at their home.   The picture I include here is from that visit.  Jan and I finally talked about what was on my mind in March, 2023 when I visited him in Leiden Hospital, quantum gravity.  It turns out we were thinking about the same problem: Penrose's demonstration that the gravitational self energy places restrictions on the time scale for a quantum superposition of two spatially separated objects.   One of his colleagues at Leiden, Tjerk Oosterkamp\cite{to}, is performing experiments to test Penrose's ideas.  The synergy with Jan never ceased to surprise me.  Our conversations were always endless.  It turns out Jan was writing a book about this very problem while I was trying to link Penrose to a recent proposal by Cotler/Strominger\cite{cs} in which unitarity is abandoned.  Jan's book, to be published soon by Oxford Press, is entitled ironically ``On Time.''  That's exactly what he ran out of.  Time.   His mark on us all is indelible and his life way too short.  We will all miss Jan but his loss is harshest for his family, Christa Buschmann, his widow, his son, Filip, his sister, Mieke, and his favorite niece, Amber.
  
\bibliographystyle{IEEEtran}
\bibliography{z}

\begin{thebibliography}{1}
\providecommand{\url}[1]{#1}
\csname url@samestyle\endcsname
\providecommand{\newblock}{\relax}
\providecommand{\bibinfo}[2]{#2}
\providecommand{\BIBentrySTDinterwordspacing}{\spaceskip=0pt\relax}
\providecommand{\BIBentryALTinterwordstretchfactor}{4}
\providecommand{\BIBentryALTinterwordspacing}{\spaceskip=\fontdimen2\font plus
\BIBentryALTinterwordstretchfactor\fontdimen3\font minus \fontdimen4\font\relax}
\providecommand{\BIBforeignlanguage}[2]{{%
\expandafter\ifx\csname l@#1\endcsname\relax
\typeout{** WARNING: IEEEtran.bst: No hyphenation pattern has been}%
\typeout{** loaded for the language `#1'. Using the pattern for}%
\typeout{** the default language instead.}%
\else
\language=\csname l@#1\endcsname
\fi
#2}}
\providecommand{\BIBdecl}{\relax}
\BIBdecl

\bibitem{hf}
\BIBentryALTinterwordspacing
J.~Zaanen and O.~Gunnarsson, ``Charged magnetic domain lines and the magnetism of high-${T}_{c}$ oxides,'' \emph{Phys. Rev. B}, vol.~40, pp. 7391--7394, Oct 1989. [Online]. Available: \url{https://link.aps.org/doi/10.1103/PhysRevB.40.7391}
\BIBentrySTDinterwordspacing

\bibitem{ldau}
\BIBentryALTinterwordspacing
V.~I. Anisimov, J.~Zaanen, and O.~K. Andersen, ``Band theory and mott insulators: Hubbard u instead of stoner i,'' \emph{Phys. Rev. B}, vol.~44, pp. 943--954, Jul 1991. [Online]. Available: \url{https://link.aps.org/doi/10.1103/PhysRevB.44.943}
\BIBentrySTDinterwordspacing

\bibitem{ads}
\BIBentryALTinterwordspacing
M.~{\v{C}}ubrovi{\'c}, J.~Zaanen, and K.~Schalm, ``String theory, quantum phase transitions, and the emergent fermi liquid,'' \emph{Science}, vol. 325, no. 5939, pp. 439--444, 2009. [Online]. Available: \url{https://www.science.org/doi/abs/10.1126/science.1174962}
\BIBentrySTDinterwordspacing

\bibitem{sawa}
\BIBentryALTinterwordspacing
M.~B.~J. Meinders, H.~Eskes, and G.~A. Sawatzky, ``Spectral-weight transfer: Breakdown of low-energy-scale sum rules in correlated systems,'' \emph{Phys. Rev. B}, vol.~48, pp. 3916--3926, Aug 1993. [Online]. Available: \url{https://link.aps.org/doi/10.1103/PhysRevB.48.3916}
\BIBentrySTDinterwordspacing

\bibitem{planckian}
\BIBentryALTinterwordspacing
J.~Zaanen, ``Why the temperature is high,'' \emph{Nature}, vol. 430, no. 6999, pp. 512--513, 2004. [Online]. Available: \url{https://doi.org/10.1038/430512a}
\BIBentrySTDinterwordspacing

\bibitem{zsa}
\BIBentryALTinterwordspacing
J.~Zaanen, G.~A. Sawatzky, and J.~W. Allen, ``Band gaps and electronic structure of transition-metal compounds,'' \emph{Phys. Rev. Lett.}, vol.~55, pp. 418--421, Jul 1985. [Online]. Available: \url{https://link.aps.org/doi/10.1103/PhysRevLett.55.418}
\BIBentrySTDinterwordspacing

\bibitem{to}
\BIBentryALTinterwordspacing
J.~van Wezel, T.~Oosterkamp, and J.~Zaanen, ``Towards an experimental test of gravity-induced quantum state reduction,'' \emph{Philosophical Magazine}, vol.~88, no.~7, pp. 1005--1026, 2008. [Online]. Available: \url{https://doi.org/10.1080/14786430801941824}
\BIBentrySTDinterwordspacing

\bibitem{cs}
J.~Cotler and A.~Strominger, ``The universe as a quantum encoder,'' \emph{arXiv preprint arXiv:2201.11658}, 2022.

\end{thebibliography}
  \end{document}